\documentclass[12pt]{article}
\usepackage{epsf,epsfig,amssymb,amsmath,graphicx,float}

\newcommand{\lsim}{\raisebox{-0.13cm}{~\shortstack{$<$ \\[-0.07cm] $\sim$}}~}

\begin{document}
\renewcommand{\thefootnote}{\fnsymbol{footnote}}

\begin{titlepage}

\begin{center}

\vspace{1cm}

{\Large {\bf Thermal abundance of non-relativistic relics with
    Sommerfeld enhancement }}

\vspace{1cm}

{\bf Hoernisa Iminniyaz}$^{a,\,b}$\footnote{wrns@xju.edu.cn},
{\bf Mitsuru Kakizaki}$^{c\,}$\footnote{kakizaki@lapp.in2p3.fr} \\

\vskip 0.15in
{\it
$^a${Center for High Energy Physics, Peking University, Peking, 100871, China} 
\\$^b${School of Physics Science and Technology, Xinjiang University, Urumqi,
  830046, China}\\
$^c${LAPTH, Universit\'e de Savoie, CNRS, B.P. 110, F-74941 Annecy-le-Vieux Cedex, France}
}
\vskip 0.5in

\abstract{We propose an analytic treatment for computing the relic
  abundances of non-relativistic particles whose annihilation rate at
  chemical decoupling is increased by Sommerfeld enhancement.  We find
  approximate rational functions that closely fit the thermal average
  of Sommerfeld-enhanced cross sections in the massless limit of
  force carriers for $s$- and $p$-wave annihilations.  We demonstrate
  that, with the approximate thermally-averaged cross sections
  implemented, the standard analytic method for the final relic
  abundances provides accuracy to within $1\%$ even for the case of
  Sommerfeld enhancement.}

\end{center}
\end{titlepage}
\setcounter{footnote}{0}

\section{Introduction}

The determination of the relic abundance of particles which decouple
from thermal equilibrium in the early universe is indispensable for
understanding the history of the universe.  Important examples include
the computation of the cosmological dark matter abundance, which
provides us a crucial hint for screening dark matter candidates and
models beyond the standard model (SM) of particle physics, as well as
cosmological scenarios in the early universe.  Among many particles
proposed, stable or long-lived weakly interacting massive particles
(WIMPs) with weak-scale mass are excellent candidates because,
assuming the thermal production scenario, the predicted relic density
coincides with the dark matter density \cite{kotu,review}.  The value
extracted from the Wilkinson Microwave Anisotropy Probe (WMAP) data is
\cite{wmap},
\begin{eqnarray} \label{wmap}
  \Omega_{\rm DM} h^2 = 0.1109 \pm 0.0056\, ,
\end{eqnarray}
where $\Omega_{\rm DM}$ is the ratio of the dark matter mass density
to the critical density, and $h = 0.710 \pm 0.025$ is the scaled
Hubble parameter in units of 100 km sec$^{-1}$ Mpc$^{-1}$.  The
uncertainty will be improved by the PLANCK satellite \cite{planck}.

In order to calculate the number density of relic particles $\chi$
accurately, in principle one must solve the Boltzmann equation, which
describes the evolution of the distribution function.  In the standard
thermal WIMP production scenario, where decoupling occurs in the
radiation-dominated epoch, the particle number density is determined
only by the thermal average of the product of the annihilation cross
section $\sigma$ and the relative velocity of the annihilating
particles $v$.  In many cases, the thermally-averaged annihilation
cross section can be expanded in a power series: $\langle \sigma v
\rangle = a + 6b/x + \cdots$, where $x$ is the ratio of 
the particle's mass $m_\chi$ to the temperature $T$, leading to simple
analytic formulas for the final abundance \cite{standard,improved},
although there are some exceptional cases \cite{except}.  The desired
cross section for reconciling with the WMAP range is found to be
approximately $\langle \sigma v \rangle \sim 3 \times 10^{-26}$ cm$^3$
sec$^{-1}$ for the temperature at which WIMPs decouple from the
thermal bath.  Analytic methods for the abundance of relic particles
have been developed also in various non-standard cosmological
scenarios where the relic abundance is increased or decreased due to a
low reheat temperature, the late decay of a scalar field, entropy
production at late times, modification of the Hubble expansion rate,
or their combination
\cite{entropy,modified,inflaton,wino,lowtr,inflaton2,kination,quintessence,decay,st,xd,dk,ssc}.
Such analytic approaches enable us to estimate the relic abundance
without tedious numerical computations.

Recently, it was pointed out that the annihilation rate of dark matter
particles can be significantly altered by the so-called Sommerfeld
enhancement at low velocities
\cite{Sommerfeld,Hisano,Hisano_relic,Cirelli}.  Anomalous excesses of
cosmic positrons reported by PAMELA \cite{PAMELA}, ATIC \cite{ATIC}
and FERMI \cite{Fermi} have motivated us to investigate the Sommerfeld
effect because the resulting annihilation cross section is extremely
boosted for lower velocities \cite{Cirelli2,ArkaniHamed,Pospelov}.
However, it should be also emphasized that the relic abundance can be
significantly reduced by the Sommerfeld effect at chemical decoupling
\cite{Hisano_relic,Cirelli}.  There is also a possibility of a change
in the relic abundance after kinetic decoupling due to chemical
recoupling of the annihilation interactions \cite{Feng}.  
(Semi-)analytic treatment for relic abundances for the case of
Sommerfeld-enhanced $s$-wave annihilations has been discussed
\cite{Kamionkowski,Dent,Zavala,Feng2009}.
It was shown that to a very good approximation the standard analytic
method reproduces numerically computed relic abundances for
Sommerfeld-enhanced $s$-wave annihilations in the limit $\alpha/v \gg
1$, where $\alpha$ is a coupling constant between the WIMP and force
carrier \cite{Dent}.
However until now, no analytic formula for approximating relic
abundances has been discussed for arbitrary $\alpha/v$, even for the
case of massless force carriers.

In this paper, we address the relic abundance of non-relativistic
particles whose annihilation rate is altered by the Sommerfeld
enhancement when the relic particles decouple from the thermal
background\footnote{In Ref. \cite{Feng} the case with massive force carriers 
is discussed.}. We find highly accurate functions that describes the 
transition from non-enhanced thermally averaged cross sections to 
$1/v$-enhancement not only for $s$-wave but also for $p$-wave annihilations.
We then show that the standard analytic method for the final relic
abundances provides accuracy to within $1\%$ even in the range where
$1/v$ approximation does not work.  The derived approximate formula is
a powerful tool for estimating relic abundances before potential
chemical recoupling.

This paper is arranged as follows. In Section 2, we discuss our method
for approximating the thermal average of Sommerfeld-enhanced
annihilation cross sections.  In Section 3, we describe the standard
method for deriving the relic density including Sommerfeld
enhancement, and compare our analytic results to numerical
computations.  Section 4 is devoted to our conclusions.

\section{Thermally-averaged annihilation cross section}

In this section, we discuss approximated expressions for the thermal
average of Sommerfeld-enhanced WIMP annihilation cross sections.

When WIMPs decouple from thermal background, they are
non-relativistic.  In the absence of force carrier, the annihilation
cross section of WIMPs can be expanded with respect to the relative
velocity $v$,
\begin{eqnarray}
  \sigma_0 v = a + b v^2 + {\cal O}(v^4)\, ,
\end{eqnarray}
where $a,b$ are constants.  For $s$-wave annihilation, $a$ gives the
dominant contribution to the annihilation of WIMPs.  If the $s$-wave
contribution is suppressed, $b$ is described by the $p$-wave
contribution.  When massless force carriers mediate interactions
between annihilating particles, the annihilation cross section is
enhanced by the factor \cite{Iengo},
\begin{eqnarray}
  S_l = \left[ \prod_{s=1}^{l}
    \left(s^2 + \frac{\alpha^2}{v^2} \right) \right] {\rm e}^{\pi
    \alpha/v} \frac{\pi \alpha/v}{\sinh (\pi \alpha/v) (l!)^2}\, ,
\end{eqnarray}
for $l$-partial wave.  Here, $\alpha$ is a coupling strength.  For
example, for the case where the annihilation rate of a
fermion-antifermion pair is enhanced by scalar boson exchanges, the
coupling strength is given by $\alpha=f^2/(4 \pi)$, with $f$ being the
Yukawa coupling constant.  In this paper, we focus on the first two
lowest modes,
\begin{eqnarray} \label{eq:sommer}
  S_s = \frac{2 \pi \alpha/v}{1- {\rm e}^{- 2 \pi \alpha/v}}\, ,
  \quad 
  S_p = \left(1 + \frac{\alpha^2}{v^2} \right)
  \frac{2 \pi \alpha/v}{1- {\rm e}^{- 2 \pi \alpha/v}}\, .
\end{eqnarray}
The case of $\alpha/v \ll 1$ results in the usual non-enhanced
annihilation cross sections as $S_l$ approaches to unity.  In the
opposite limit $\alpha/v \gg 1$, the Sommerfeld enhancement factor is
simplified down to $S_s = 2 \pi \alpha/v$ for $s$-wave annihilation,
and to $S_p = 2 \pi \alpha^3/v^3$ for $p$-wave annihilation.  At the
leading order of the $v^2$ expansion, we can parametrize the
Sommerfeld-enhanced annihilation cross sections as
\begin{eqnarray} \label{eq:sigmav}
  \sigma v  =  \left\{ 
    \begin{array}{ll}
      a S_s & (s$-${\rm wave}) \\
      b v^2 S_p & (p$-${\rm wave})\, .
    \end{array}
  \right. 
\end{eqnarray}
Inclusion of higher order terms is a trivial extension of the
procedure we will present.

Let us consider the enhancement of the WIMP annihilation by
introducing the boost factor $B = \langle \sigma v \rangle/\langle
\sigma_0 v \rangle$.  The thermal average of the $s$-wave annihilation
cross section is given by
\begin{eqnarray}
  \langle \sigma v \rangle = a \langle S_s \rangle
  = a ~ \frac{x^{3/2}}{2 \sqrt{\pi}}  \int_0^{\infty} \!\! {\rm d}v~
  v^2~ {\rm e}^{-x v^2/4}~ \frac{2 \pi \alpha/v}{1- {\rm e}^{- 2 \pi \alpha/v}}\, .
\end{eqnarray}
Introducing further the variables $y \equiv \alpha
\sqrt{\pi x}$ and $t \equiv v \sqrt{x}/(2 \sqrt{\pi})$, the boost
factor is solely described by $y$,
\begin{eqnarray} \label{eq:bs}
  B_s(y) = \langle S_s \rangle = 4 \pi y \int_0^{\infty} \!\!
  {\rm d} t~t~{\rm e}^{-\pi t^2}~\frac{1}{1 - {\rm e}^{-y/t}}\, .
\end{eqnarray}
Similarly, for $p$-wave annihilation, we obtain
\begin{eqnarray}  \label{eq:bp}
  B_p(y) = \frac{x \langle v^2 S_p \rangle}{6}
  = \frac{8 \pi^2 y}{3}  \int_0^{\infty} \!\! {\rm d}t~
  t^3~ {\rm e}^{-\pi t^2}~ \left(1 + \frac{y^2}{4 \pi^2 t^2} \right) 
  \frac{1}{1- {\rm e}^{- y/t}}\, .
\end{eqnarray}

In the case where the coupling constant $\alpha$ is small or the
temperature is high enough to suppress Sommerfeld enhancement, we can
expand the boost factor in the Taylor series,
\begin{eqnarray} \label{eq:btaylor}
  B_{{\rm Taylor},s}(y) & = & 1 + y + \frac{\pi}{6} y^2\, ,
  \nonumber \\
  B_{{\rm Taylor},p}(y) & = & 
  1 + \frac{2}{3}y + \frac{3 + \pi^2}{18 \pi} y^2\, .
\end{eqnarray}
In the opposite limit, where the cross section is enhanced by $1/v$,
${\rm e}^{-y/t}$ in the denominators of Eqs.(\ref{eq:bs}) and
(\ref{eq:bp}) are negligible, leading to
\begin{eqnarray} \label{eq:vinv}
  B_{1/v,s}(y) = 2 y\, , \quad 
  B_{1/v,p}(y) = \frac{y^3}{3 \pi} + \frac{4}{3}y\, .
\end{eqnarray}
As a simple function that connects the two limiting results for
$y\to0$ and $y \to \infty$, we propose the following interpolations:
\begin{eqnarray} \label{eq:bapp}
  B_{{\rm app},s}(y) & = & \frac{ 1 + 7y/4 + 3y^2/2
    + (3/2 - \pi/3) y^3 }
  { 1+ 3y/4 + (3/4 - \pi/6)y^2}\, ,
  \nonumber \\
  B_{{\rm app},p}(y) & = & \frac{ 1 + 11y/12 + (1/(6 \pi) + 1/6 + \pi/18)y^2
    + y^3/(3 \pi) + y^4/(12 \pi) }{1 + y/4}\, .
\end{eqnarray}
Notice that these choices are not unique.  We found that the above
expressions are ones of the simplest fitting functions that can
reproduce the exact numerical results for the whole range of $y$, as
we will see below.

Figure \ref{fig:b} compares various approximations against the exact
boost factor.  In Fig.\ref{fig:b}(a) (Fig.\ref{fig:b}(b)), the exact
boost factor $B_{s(p)}$, Eq.(\ref{eq:bs}) (Eq.(\ref{eq:bp})) (solid
line), its Taylor series up to the quadratic order $B_{{\rm
    Taylor},s(p)}$, Eq.(\ref{eq:btaylor}) (dashed), $1/v$
approximation $B_{1/v,s(p)}$, Eq.(\ref{eq:vinv}) (dotted), and our
approximation $B_{{\rm app},s(p)}$, Eq.(\ref{eq:bapp}) (+),
are shown as a function of $y = \alpha \sqrt{\pi x}$.  Notice that our
approximation $B_{{\rm app},s(p)}$ completely falls together with its
exact results.  For the case of $s$-wave ($p$-wave) annihilation, the
accuracy of the Taylor expansion Eq.(\ref{eq:btaylor}) decreases down
to $99 \%$ for $y = 0.45$ ($y = 0.64$); the range for the $1/v$
approximation Eq.(\ref{eq:vinv}) to work at this level is $y > 3.0$
($y > 3.6$).  The in-between range $0.45 < y < 3.0$ ($0.64 < y <
3.6$), where neither of the known approximations works, corresponds to
$0.06 \lsim \alpha \lsim 0.4$ ($0.08 \lsim \alpha \lsim 0.5$) for the
typical WIMP decoupling temperature $T \sim m_\chi/20$.  On the other
hand, our ansatz Eq.(\ref{eq:bapp}) always reproduces the exact
results with accuracy of less than $0.3 \%$ ($0.9 \%$).

\begin{figure}[t!]
  \begin{center}
    \hspace*{-0.5cm} \scalebox{0.63}{\includegraphics*{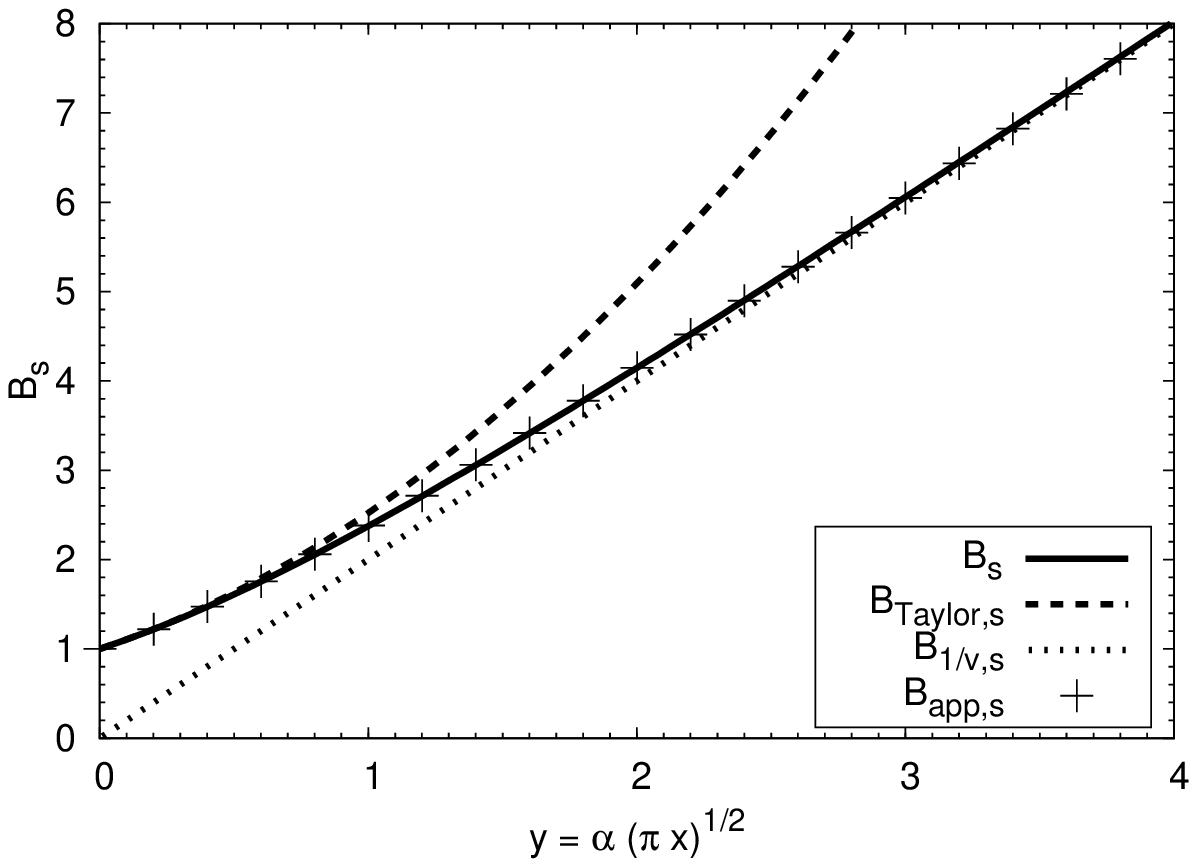}}
    \put(-115,-12){(a)} 
    \hspace*{-0.5cm} \scalebox{0.63}{\includegraphics*{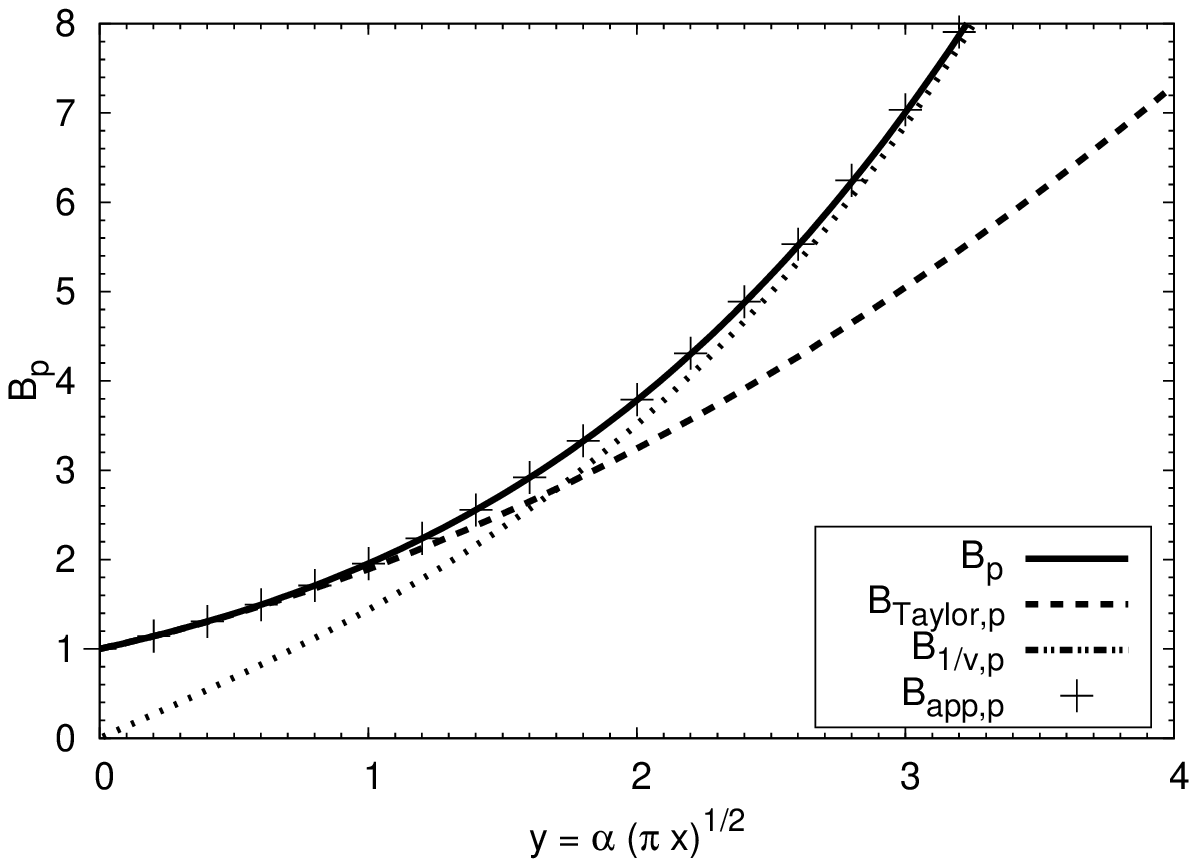}} 
    \put(-115,-12){(b)}
    \caption{\label{fig:b} \footnotesize Exact boost factor
      $B_{s(p)}$, Eq.(\ref{eq:bs}) (Eq.(\ref{eq:bp})) (solid line),
      its Taylor series up to the quadratic order $B_{{\rm
          Taylor},s(p)}$, Eq.(\ref{eq:btaylor}) (dashed), $1/v$
      approximation $B_{1/v,s(p)}$, Eq.(\ref{eq:vinv}) (dotted), and
      our approximation $B_{{\rm app},s(p)}$, Eq.(\ref{eq:bapp})
      (+), as a function of $y = \alpha \sqrt{\pi x}$.  Our
      approximation $B_{{\rm app},s(p)}$ completely falls together
      with its exact results $B_{s(p)}$. }
  \end{center}
\end{figure}

\section{Relic Abundance with Sommerfeld Enhancement}

Let us discuss the computation of the relic abundance for the case
where the annihilation cross section is enhanced by the Sommerfeld
factor Eq.(\ref{eq:sommer}). As an analytic formalism for the
computation of the relic abundance, we follow the standard freeze-out
picture \cite{kotu,standard}, with appropriate modifications
\cite{Feng,Kamionkowski,Dent,Zavala,Feng2009}. Here, we show that the standard
formalism works even for the case of Sommerfeld enhancement.

The relic density of thermal relic particles $\chi$ whose single
production and decay are forbidden by some symmetry is determined by
solving the Boltzmann equation,
\begin{eqnarray} \label{eq:boltzmann_n}
  \frac{{\rm d}n_{\chi}}{{\rm d}t} + 3 H n_{\chi} =  - \langle \sigma v\rangle
  (n^2_{\chi} - n_{\chi,{\rm eq}}^2)~,
\end{eqnarray}
which describes the time evolution of the number density $n_{\chi}$ of
the $\chi$ particles in the universe expanding at the rate $H$.  Here
$n_{\chi,{\rm eq}}$ is the equilibrium value of $n_\chi$, whose
non-relativistic limit is given by
\begin{equation} \label{eq:eq}
  n_{\chi,{\rm eq}} = g_\chi ~{\left( \frac{m_\chi T}{2 \pi} \right)}^{3/2}
  {\rm e}^{-m_\chi/T}~,
\end{equation}
where $g_{\chi}$ denotes the internal degrees of freedom of the $\chi$
particle.  At high temperatures, the $\chi$ particles are in thermal
equilibrium. After $T$ drops below $m_\chi$, the number density
$n_\chi$ exponentially decreases until the interaction rate $\Gamma =
n_\chi \langle \sigma v \rangle$ falls below the expansion rate in the
radiation-dominated epoch $H = (\pi T^2/M_{\rm Pl}) \sqrt{g_*/90}$,
where $M_{\rm Pl} = 2.4\times 10^{18}$ GeV is the reduced Planck mass,
and $g_*$ the number of the relativistic degrees of freedom. Then, the
$\chi$ particles are no longer kept in thermal equilibrium and the
comoving number density becomes fixed.

We can express the Boltzmann equation (\ref{eq:boltzmann_n}) in terms
of the dimensionless quantities $Y_\chi = n_\chi/s$ and $x =
m_\chi/T$, where $s = (2 \pi^2/45) g_* T^3$ is the entropy density.
Assuming that the universe expands adiabatically, the Boltzmann
equation can be rewritten
\begin{eqnarray} \label{eq:boltzmanns} \frac{{\rm d} Y}{{\rm d}x} = -
  \frac{4 \pi}{\sqrt{90}} m_\chi M_{\rm Pl}
  \frac{\sqrt{g_*} \langle \sigma v \rangle}{x^2}
  (Y_\chi^2 - Y_{\chi,{\rm eq}}^2)\, .
\end{eqnarray}
Introducing the variable $\Delta = Y_{\chi} -Y_{\chi, {\rm eq}}$, we obtain 
\begin{equation} \label{eq:delta} \frac{{\rm d} \Delta}{{\rm d}x} = -
  \frac{{\rm d} Y_{\chi, {\rm eq}}}{{\rm d}x} - \frac{4
    \pi}{\sqrt{90}} m_\chi M_{\rm Pl} \frac{\sqrt{g_*} \langle \sigma v \rangle
    }{x^2}~ \Delta (2 Y_{\chi,{\rm eq}} +
  \Delta)\, .
\end{equation}
The solution can be analytically derived in two extreme regimes.  At
temperatures above the freeze-out temperature $T_F$, the
$\chi$ particles are in thermal equilibrium, so that the deviation of
$Y_{\chi}$ from its equilibrium value $Y_{\chi, {\rm eq}}$ is very
small.  Ignoring $\Delta^2$ and ${\rm d} \Delta/{\rm d}x$, the
solution is given by
\begin{eqnarray} \label{eq:early}
  \Delta \simeq \frac{x^2}{(8 \pi/\sqrt{90}) m_\chi M_{\rm Pl}
  \sqrt{g_*} \langle \sigma v \rangle}\, ,
\end{eqnarray}
where we have used ${\rm d} Y_{\chi,{\rm eq}}/{\rm d}x \simeq -
Y_{\chi,{\rm eq}}$ for $x \gg 1$.  Freeze-out occurs when $Y_\chi$
deviates from $Y_{\chi, {\rm eq}}$:
\begin{equation}
  \Delta (x_F) = c Y_{\chi,{\rm eq}}(x_F)\, ,
\end{equation}
where $x_F = m_\chi/T_F$, and $c$ is a numerical constant of order of
unity.  Using the early time solution, Eq.(\ref{eq:early}), we
obtain the recursive equation for determining the value of $x_F$,
\begin{eqnarray}
  x_F = \left. \ln \left( \sqrt{\frac{45}{\pi^5}} c m_\chi M_{\rm Pl} g_\chi
    \frac{\langle \sigma v \rangle}{\sqrt{x g_*}}
    \right) \right|_{x=x_F}\, ,
\end{eqnarray}
For the standard $s$- and $p$-wave annihilation cross sections, the
choice of $c=\sqrt{2} -1$ is known to give a good agreement with the
numerical results.  We will see that this choice is still valid even
with Sommerfeld enhancement.  At temperatures below $T_F$, the
production term $Y_{\chi,{\rm eq}}$ in the Boltzmann equation can be
ignored.  Therefore, the final relic abundance is found to be
\begin{eqnarray}
  Y_\chi(x \to \infty)
  = \frac{1}{(4 \pi/\sqrt{90}) m_\chi M_{\rm Pl} I (x_F)}\, ,
\end{eqnarray}
where the annihilation integral is defined by
\begin{eqnarray}
  I(x_F) &=& \int^{\infty}_{x_F}
  \!\! {\rm d}x \frac{g_* \langle \sigma v \rangle}{x^2}\, .
\end{eqnarray}
For convenience, we express the final abundance in terms of
$\Omega_\chi h^2 =m_\chi s_0 Y_{\chi}(x \to \infty) h^2/\rho_{\rm
  crit}$, where $s_0 = 2.9 \times 10^3~{\rm cm}^{-3}$ is the present
entropy density, and $\rho_{\rm crit} = 3 M_{\rm Pl}^2 H^2$ is the
critical density.  The conversion factor of the annihilation integral
to the relic density is given by
\begin{eqnarray}
  \Omega_\chi h^2 = \frac{8.5 \times 10^{-11}}{I(x_F)~{\rm GeV}^2}\, .
\end{eqnarray}
In sharp contrast to the exact boost factors, Eqs.(\ref{eq:bs}) and
(\ref{eq:bp}), our ansatz $B_{\rm app}$,
Eq.(\ref{eq:bapp}), leads to simple analytic annihilation integrals.
For $s$-wave annihilation,
\begin{eqnarray} \label{eq:integral_s}
  \frac{I(x_F)}{a \sqrt{g_*}} &=& \int^{\infty}_{x_F}
  \!\! {\rm d}x \frac{B_{{\rm app},s}}{x^2}
  \nonumber \\
  & = &
  \frac{1}{x_F} + 2 \alpha \sqrt{\frac{\pi}{x_F}}
 +  \frac{\pi^2 \alpha^2}{6} {\rm ln} \left( 1
   + \frac{9 \alpha \sqrt{\pi x_F} + 12}{(9 - 2 \pi) \pi \alpha^2 x_F} \right)
  \nonumber \\
  &&+ \pi \alpha^2 \frac{36 - 11 \pi}{\sqrt{3(117-32 \pi)}}
  \left( \frac{\pi}{2} - \tan^{-1} {\frac{2  (9 - 2 \pi) \alpha \sqrt{\pi x_F} + 9}
  {\sqrt{3(117 - 32 \pi)}}}  \right)\, .
\end{eqnarray}
For $p$-wave annihilation,
\begin{eqnarray} \label{eq:integral_p}
  \frac{I(x_F)}{6b \sqrt{g_*}} &=& \int^{\infty}_{x_F}
  \!\! {\rm d}x \frac{B_{{\rm app},p}}{x^3}
  \nonumber \\
  & = &
  \frac{1}{2 x_F^2} + \frac{4 \alpha \sqrt{\pi}}{9 x_F^{3/2}}
  + \frac{(3 + \pi^2) \alpha^2}{18 x_F} 
  + \frac{(21 - \pi^2) \sqrt{\pi} \alpha^3}{36 \sqrt{x_F}} 
  \nonumber \\
  && + \frac{\pi (3 + \pi^2) \alpha^4}{144}
  {\rm ln} \left( 1 + \frac{4}{\alpha \sqrt{\pi x_F} }\right)\, .
\end{eqnarray}

\begin{figure}[t!]
  \begin{center}
    \hspace*{-0.5cm}
    \scalebox{0.63}{\includegraphics*{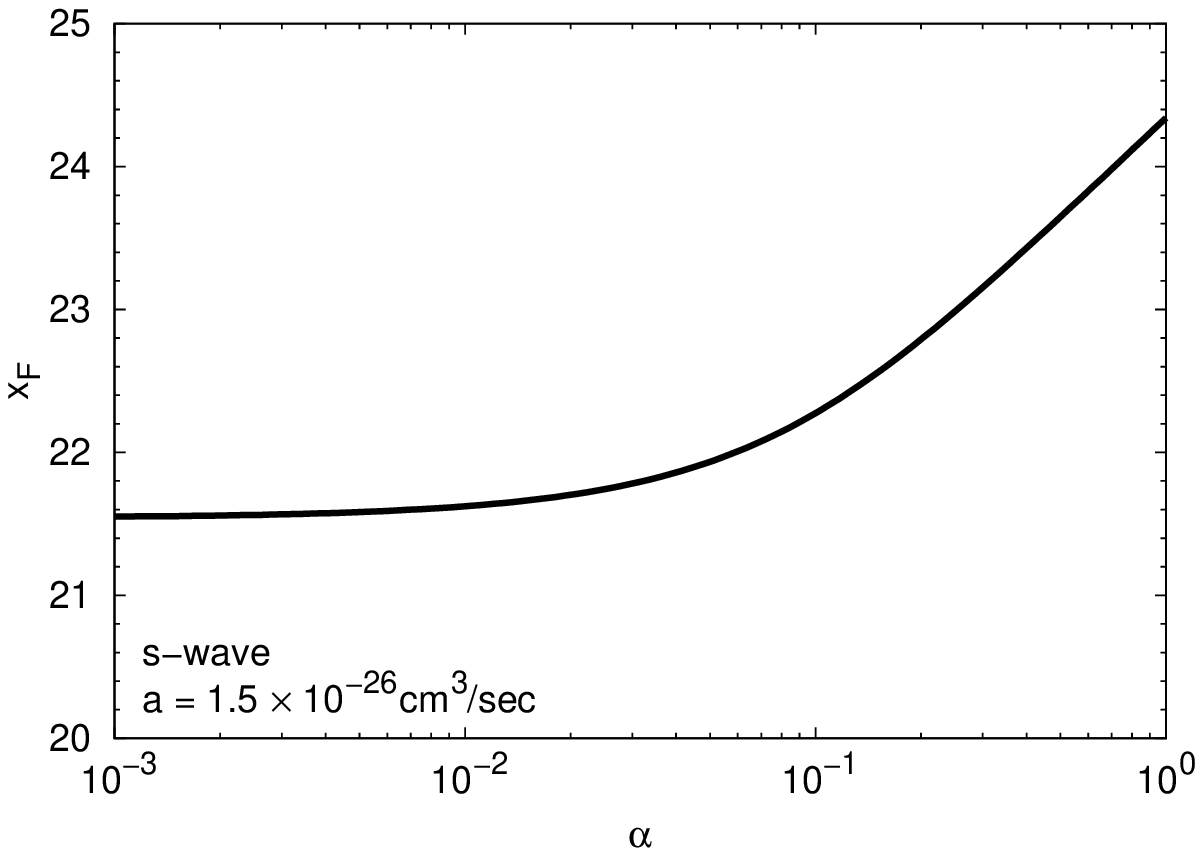}}
    \put(-115,-12){(a)}
    \scalebox{0.63}{\includegraphics*{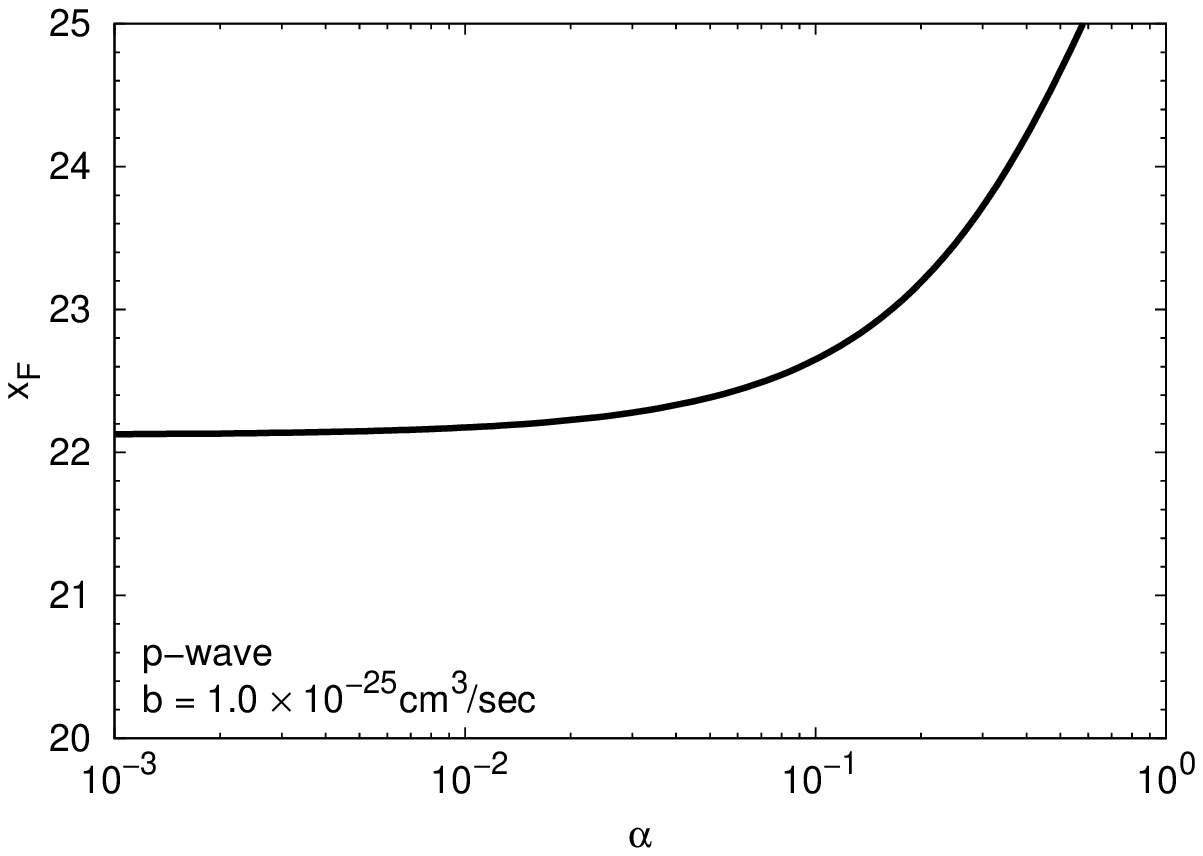}}
    \put(-115,-12){(b)}
    \caption{\label{fig:xfsom} \footnotesize Normalized inverse
      freeze-out temperature as a function of $\alpha$ for the
      Sommerfeld-enhanced $s$-wave annihilation with $a = 1.5 \times
      10^{-26}~{\rm cm}^3/{\rm sec}$ (a), and $p$-wave annihilation
      with $b = 1.0 \times 10^{-25}~{\rm cm}^3/{\rm sec}$ (b).  Here
      we take $m_\chi =100$ GeV, $g_{\chi} = 2$ and $g_* = 90$.}
  \end{center}
\end{figure}

In Fig.\ref{fig:xfsom}, we show the normalized inverse freeze-out
temperature $x_F$ as a function of the coupling constant $\alpha$ for
the Sommerfeld-enhanced $s$-wave annihilation with $a = 1.5 \times
10^{-26}~{\rm cm}^3/{\rm sec}$ (a), and $p$-wave annihilation with $b
= 1.0 \times 10^{-25}~{\rm cm}^3/{\rm sec}$ (b).  Here we take
$m_\chi=100$ GeV, $g_{\chi} = 2$ and $g_* = 90$.  The plots illustrate
that the freeze-out temperature decreases significantly as the
coupling constant $\alpha$ increases.

\begin{figure}[t!]
  \begin{center}
    \hspace*{-0.5cm}
    \scalebox{0.63}{\includegraphics*{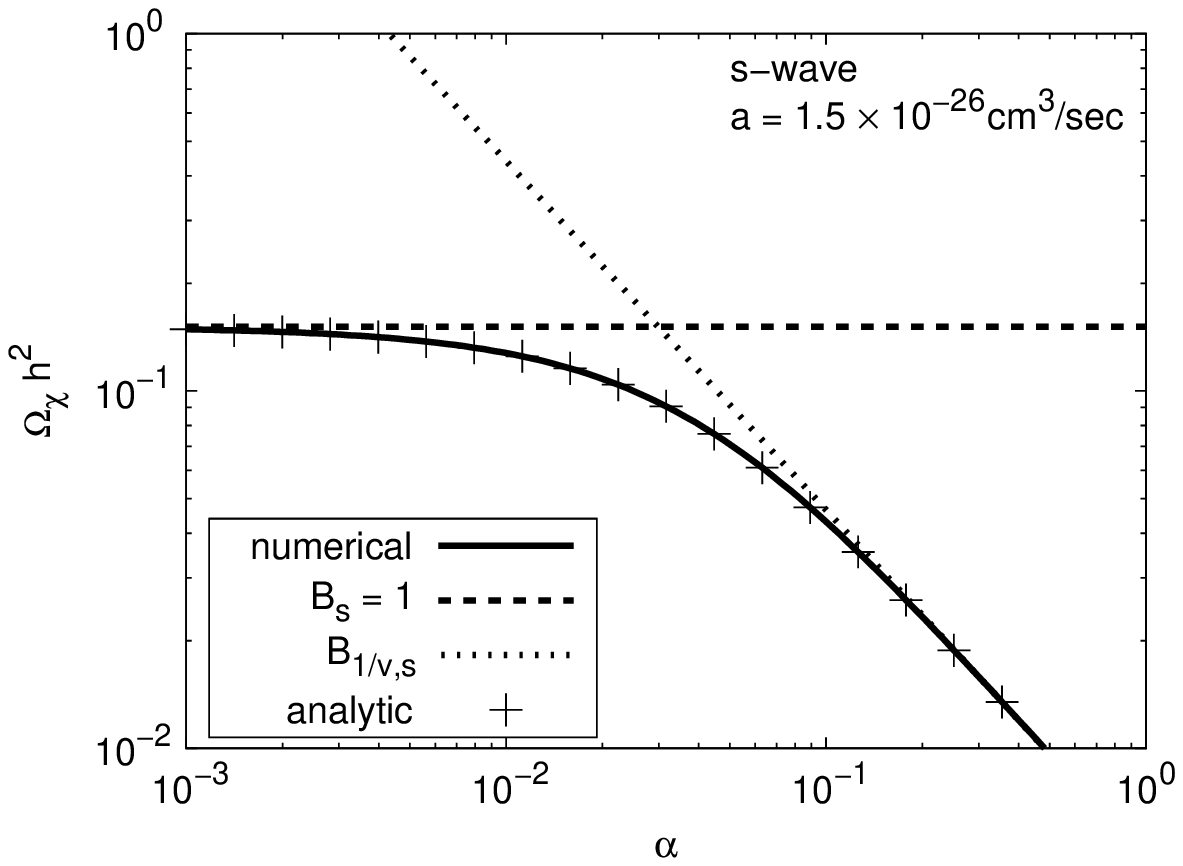}}
    \put(-115,-12){(a)}
    \scalebox{0.63}{\includegraphics*{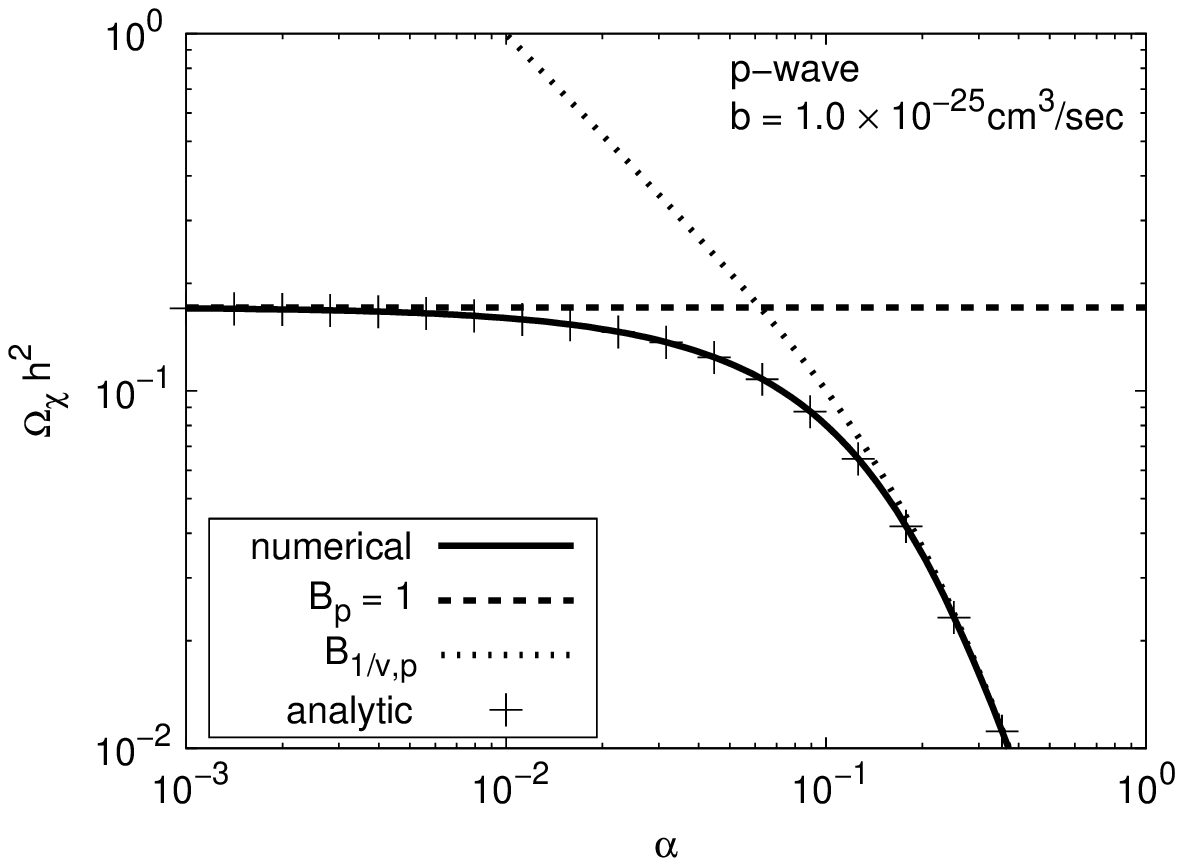}}
    \put(-115,-12){(b)}
    \caption{\label{fig:omegah2} \footnotesize Comparison of the
      analytically estimated relic abundance (+) with the numerically
      computed result (solid line).  Here we take the same parameter
      set as in Fig.\ref{fig:xfsom}.  The cases of $B_s = 1$ ($B_p = 1$)
      (dashed) and $1/v$-enhancement (dotted) are also shown.  Notice
      that the analytically computed relic abundance using
      Eq.(\ref{eq:integral_s}) or (\ref{eq:integral_p}) (+) completely
      falls together with the numerical result (solid).}
  \end{center}
\end{figure}

Figure \ref{fig:omegah2}(a) compares the analytically estimated relic
abundance using Eq.(\ref{eq:integral_s}) (+) with the numerically
computed result (solid line) for the same parameter set as in
Fig.\ref{fig:xfsom}(a).  For the $p$-wave case, the approximated relic
abundance using Eq.(\ref{eq:integral_p}) (+) and its exact result
(solid) are displayed in Fig.\ref{fig:omegah2}(b).  We also show the
relic abundances for $B_s = 1$ ($B_p = 1$) (dashed) and those for the
$1/v$-enhancement case (dotted).  We emphasize that our analytic
result successfully describes the transition from the non-enhanced
case, $B_s = 1$ ($B_p = 1$) , to $1/v$-enhancement, and overlaps with the
numerically computed results even in the intermediate region $3 \times
10^{-3} \lsim \alpha \lsim 0.1$, where neither of the limiting
approximations works.  In other words, the standard analytic method
for the final abundance is also applicable to the Sommerfeld-enhanced
annihilation cross section, which exhibits a non-trivial velocity
dependence.  The deviation of the analytically derived relic abundance
from the numerical result is found to be always less than $1\%$.

\section{Conclusions}

In this paper, we have proposed an approximate analytic function for
evaluating the boost factor caused by either $s$- or $p$-wave
Sommerfeld-enhanced annihilation, Eq.(\ref{eq:sigmav}), assuming
massless force carriers.  For $s$-wave ($p$-wave) annihilation, our
ansatz for the boost factor smoothly connects the limiting cases of
$B_s = 1$ ($B_p = 1$) and $1/v$-enhancement, and reproduces the exact
values with accuracy of less than $0.3 \%$ ($0.9 \%$) in the whole
range of the coupling constant $\alpha$.  We have applied the
approximate boost factor $B_{{\rm app},s}$ ($B_{{\rm app},p}$) to the
computation of the thermal relic abundance of particles whose $s$-wave
($p$-wave) annihilation is Sommerfeld-enhanced in the early universe,
and derived a totally analytic formula for the final relic abundance
in terms of the freeze-out temperature $x_F$.  Our findings show that
the standard analytic method for the relic abundance attains accuracy
of less than $1\%$.  Our results are rather generic, and applicable
not only to dark matter candidates but also to other particles that
decouple non-relativistically from the thermal background in the
early universe.

Finally, we mention that chemical recoupling after kinetic decoupling
could affect the relic abundance \cite{Feng}.  A more detailed
discussion including the evolution after kinetic decoupling
will appear elsewhere.\\

After completion of this work, we received a preprint \cite{Hannestad}
that deals with a similar subject.

\section*{Acknowledgements}

The work of M.K. was partially supported by the Marie Curie Training
Research ``HEPTools'' under contract No. MRTN-CT-2006-035505. The
work of H.I. is supported by the National Natural Science
Foundation of China (11047009) and by the doctor fund BS100108 of
Xinjiang university.  H.I. also thanks to the hospitality and support
of LAPTH, Annecy during the completion of this work.


\begin{thebibliography}{99}

\bibitem{kotu}
E.~W.~Kolb and M.~S.~Turner, {\it The Early Universe}, Addison-Wesley
(Redwood City, CA, 1990).

\bibitem{review}
For a review, see G. Bertone, D. Hooper and J. Silk, Phys. Rep. {\bf 405},
279 (2005) [arXiv:hep-ph/0404175]; G. Jungman, M. kamionkowski, and K. Griest,
Phys. Rep. {\bf 267}, 195 (1996)

\bibitem{wmap}
WMAP Collab., D.~N.~Spergel {\it et al.},
Astrophys. J. Suppl. {\bf 148}, 175 (2003) [arXiv:astro-ph/0302209];
WMAP Collab., D.~N.~Spergel {\it et al.},
Astrophys. J. Suppl. {\bf 170}, 377 (2007) [arXiv:astro-ph/0603449];
WMAP Collab., E.~Komatsu {\it et al.},
Astrophys. J. Suppl. {\bf 180}, 330 (2009) [arXiv:0803.0547 [astro-ph]];
WMAP Collab., J.~Dunkley {\it et~al.},
Astrophys. J. Suppl. {\bf 180}, 306 (2009) [arXiv:0803.0586 [astro-ph]];
E.~Komatsu {\it et~al.}, arXiv:1001.4538 [astro-ph.CO];
N.~Jarosik {\it et al.}, arXiv:1001.4744 [astro-ph.CO].


\bibitem{planck}
Planck Science Team, ``Planck Bluebook,''
http://www.rssd.esa.int/planck (2005).


\bibitem{standard}
R.~J.~Scherrer and M.~S.~Turner, Phys.\ Rev.\  D {\bf 33}, 1585 (1986),
Erratum-ibid.\  D {\bf 34}, 3263 (1986).

\bibitem{improved}
P.~Gondolo and G.~Gelmini, Nucl.\ Phys.\  B {\bf 360}, 145 (1991).

\bibitem{except}
K. Griest and D. Seckel, Phys.\ Rev.\  D {\bf 43}, 3191 (1991).


\bibitem{entropy}
  R.~J.~Scherrer and M.~S.~Turner,
  Phys.\ Rev.\  D {\bf 31}, 681 (1985).

\bibitem{modified}
  M.~Kamionkowski and M.~S.~Turner,
  Phys.\ Rev.\  D {\bf 42}, 3310 (1990).

\bibitem{inflaton}
  D.~J.~H.~Chung, E.~W.~Kolb and A.~Riotto,
  Phys.\ Rev.\  D {\bf 60}, 063504 (1999)
  [arXiv:hep-ph/9809453].

\bibitem{wino}
  T.~Moroi and L.~Randall,
  Nucl.\ Phys.\  B {\bf 570}, 455 (2000)
  [arXiv:hep-ph/9906527].

\bibitem{lowtr}
  G.~F.~Giudice, E.~W.~Kolb and A.~Riotto,
  Phys.\ Rev.\  D {\bf 64}, 023508 (2001)
  [arXiv:hep-ph/0005123].

\bibitem{inflaton2}
  R.~Allahverdi and M.~Drees,
  Phys.\ Rev.\ Lett.\  {\bf 89}, 091302 (2002)
  [arXiv:hep-ph/0203118];
  Phys.\ Rev.\  D {\bf 66}, 063513 (2002)
  [arXiv:hep-ph/0205246].

\bibitem{kination}
  P.~Salati,
  Phys.\ Lett.\  B {\bf 571}, 121 (2003)
  [arXiv:astro-ph/0207396].

\bibitem{quintessence}
  S.~Profumo and P.~Ullio,
  JCAP {\bf 0311}, 006 (2003)
  [arXiv:hep-ph/0309220].

\bibitem{decay}
  C.~Pallis,
  Astropart.\ Phys.\  {\bf 21}, 689 (2004)
  [arXiv:hep-ph/0402033].

\bibitem{st}
  R.~Catena, N.~Fornengo, A.~Masiero, M.~Pietroni and F.~Rosati,
  Phys.\ Rev.\  D {\bf 70}, 063519 (2004)
  [arXiv:astro-ph/0403614].

\bibitem{xd}
  N.~Okada and O.~Seto,
  Phys.\ Rev.\  D {\bf 70}, 083531 (2004)
  [arXiv:hep-ph/0407092].

\bibitem{dk}
M.~Drees, H.~Iminniyaz and M.~Kakizaki,
Phys.\ Rev.\  D {\bf 73}, 123502 (2006)
[arXiv:hep-ph/0603165];
Phys.\ Rev.\  D {\bf 76}, 103524 (2007)
[arXiv:0704.1590 [hep-ph]];
M.~Drees, M.~Kakizaki and S.~Kulkarni,
Phys.\ Rev.\  D {\bf 80}, 043505 (2009)
[arXiv:0904.3046 [hep-ph]].

\bibitem{ssc}
 A.~B.~Lahanas, N.~E.~Mavromatos and D.~V.~Nanopoulos,
  Phys.\ Lett.\  B {\bf 649}, 83 (2007)
  [arXiv:hep-ph/0612152].



\bibitem{Sommerfeld}
A.~Sommerfeld, Annalen der Physik {\bf 403}, 257 (1931)

\bibitem{Hisano}
  J.~Hisano, S.~Matsumoto and M.~M.~Nojiri,
  Phys.\ Rev.\  D {\bf 67}, 075014 (2003)
  [arXiv:hep-ph/0212022];
  Phys.\ Rev.\ Lett.\  {\bf 92}, 031303 (2004)
  [arXiv:hep-ph/0307216];
  J.~Hisano, S.~Matsumoto, M.~M.~Nojiri and O.~Saito,
  Phys.\ Rev.\  D {\bf 71}, 063528 (2005)
  [arXiv:hep-ph/0412403].

\bibitem{Hisano_relic}
  J.~Hisano, S.~Matsumoto, M.~Nagai, O.~Saito and M.~Senami,
  Phys.\ Lett.\  B {\bf 646}, 34 (2007)
  [arXiv:hep-ph/0610249].

\bibitem{Cirelli}
  M.~Cirelli, A.~Strumia and M.~Tamburini,
  Nucl.\ Phys.\  B {\bf 787}, 152 (2007)
  [arXiv:0706.4071 [hep-ph]].



\bibitem{PAMELA}
  O.~Adriani {\it et al.}  [PAMELA Collaboration],
  Nature {\bf 458}, 607 (2009)
  [arXiv:0810.4995 [astro-ph]].

\bibitem{ATIC}
  J.~Chang {\it et al.},
  Nature {\bf 456}, 362 (2008).

\bibitem{Fermi}
  A.~A.~Abdo {\it et al.}  [The Fermi LAT Collaboration],
  Phys.\ Rev.\ Lett.\  {\bf 102}, 181101 (2009)
  [arXiv:0905.0025 [astro-ph.HE]].


\bibitem{Cirelli2}
 M.~Cirelli, M.~Kadastik, M.~Raidal and A.~Strumia,
  Nucl.\ Phys.\  B {\bf 813}, 1 (2009)
  [arXiv:0809.2409 [hep-ph]].

\bibitem{ArkaniHamed}
  N.~Arkani-Hamed, D.~P.~Finkbeiner, T.~R.~Slatyer and N.~Weiner,
  Phys.\ Rev.\  D.\ {\bf 79} 015014 (2009)
  [arXiv:0810.0713 [hep-ph]].

\bibitem{Pospelov}
  M.~Pospelov and A.~Ritz,
  Phys.\ Lett.\  B {\bf 671}, 391 (2009)
  [arXiv:0810.1502 [hep-ph]].


\bibitem{Feng}
  J.~L.~Feng, M.~Kaplinghat and H.~B.~Yu,
  arXiv:1005.4678 [hep-ph].

\bibitem{Kamionkowski}
  M.~Kamionkowski and S.~Profumo,
 Phys.\ Rev.\ Lett.\  {\bf 101} 261301 (2008)
  [arXiv:0810.3233 [astro-ph]].

\bibitem{Dent}
  J.~B.~Dent, S.~Dutta and R.~J.~Scherrer,
  Phys.\ Lett.\  B {\bf 687}, 275 (2010)
  [arXiv:0909.4128 [astro-ph.CO]].


\bibitem{Zavala}
  J.~Zavala, M.~Vogelsberger and S.~D.~M.~White,
  Phys.\ Rev.\  D {\bf 81}, 083502 (2010)
  [arXiv:0910.5221 [astro-ph.CO]].

\bibitem{Feng2009}
  J.~L.~Feng, M.~Kaplinghat and H.~B.~Yu,
  Phys.\ Rev.\ Lett.\  {\bf 104}, 151301 (2010)
  [arXiv:0911.0422 [hep-ph]].

\bibitem{Iengo}
  R.~Iengo,
  JHEP {\bf 0905}, 024 (2009)
  [arXiv:0902.0688 [hep-ph]];
  arXiv:0903.0317 [hep-ph].


\bibitem{Hannestad}
  S.~Hannestad and T.~Tram,
  arXiv:1008.1511 [astro-ph.CO].



\end{thebibliography}
\end{document}